\documentclass[12pt,reprint,noshowpacs,nofootinbib,notitlepage,amsmath]{revtex4-1}
\usepackage{setspace}
\allowdisplaybreaks
\usepackage{graphicx,color}
\usepackage[colorlinks=true,citecolor=blue,linkcolor=blue,urlcolor=blue]{hyperref}
\usepackage[charter]{mathdesign}
\usepackage{multirow}
\usepackage{enumerate}
\usepackage{slashed}

\DeclareSymbolFontAlphabet{\mathcal}{symbols}
\DeclareSymbolFont{symbols}{OMS}{xmdcmsy}{m}{n}
\DeclareSymbolFont{largesymbols}{OMX}{xmdcmex}{m}{n}
\SetSymbolFont{symbols}{bold}{OMS}{xmdcmsy}{b}{n}

\begin{document}
\title{\Large \color{blue}Maximal Acceleration, Reciprocity \& Nonlocality}
\author{Luca Buoninfante}
\email{ buoninfante.l.aa@m.titech.ac.jp}
\affiliation{Department of Physics, Tokyo Institute of Technology, Tokyo 152-8551, Japan}

\begin{abstract}
The existence of a fundamental scale is expected to be a key feature of quantum gravity. Many approaches take this property as a starting assumption.  
Here, instead, we take a less conventional viewpoint based on a critical inspection of both fundamental principles and  kinematic laws. We point out that rigorous arguments suggest a more urgent need to revise known theories to incorporate a fundamental acceleration scale already in flat space. The reciprocity principle can naturally do so. In addition to noticing links with string theory, we argue that the reciprocity principle implies an infinite-derivative generalization of the Einstein-Hilbert action that makes the gravitational interaction fundamentally nonlocal, thus providing a guiding principle that could lead us towards the formulation of a consistent theory of quantum gravity.
\\
\\
\textcolor{blue}{\textit{Essay received an Honorable Mention in the 2021 Gravity Research Foundation Essays Competition}}.
\end{abstract}
\maketitle

The three fundamental constants associated with the notions of relativity, quantum and gravity are the speed of light $c,$  (reduced) Planck's constant $\hbar$ and Newton's constant $G,$ respectively. By combining them together we can define the Planck length $\ell_{\rm p}=\sqrt{\hbar G/c^3},$ or the Planck mass $m_{\rm p}=\hbar/c\ell_{\rm p}.$ In general, these quantities are \textit{not} fundamental scales and are \textit{not} related to any notion of discretization. However, they acquire an important meaning when both quantum and gravitational effects are non-negligible. For instance, $\ell_{\rm p}$ is interpreted as the minimal observable length in any experiment because by probing smaller distances one would form a black hole~\cite{Garay:1994en}. Thus, $m_{\rm p}$ is expected to be the maximum mass for an elementary particle.

In a quantum-gravitational regime we can also define the following upper bound on the gravitational acceleration, known as Planck acceleration:
\begin{equation}
a_{\rm p}=\frac{Gm_{\rm p}}{\ell_{\rm p}^2}=\sqrt{\frac{c^7}{\hbar G}}=\frac{m_{\rm p}c^3}{\hbar}\simeq 5.6\times 10^{52}\,{\rm m/s^2}\,,\label{Planck-acc}
\end{equation}
which corresponds to the gravitational acceleration on the surface (event horizon) of the smallest and most concentrated object in Nature, i.e. a Planck-size black hole. 

These limitations seem to suggest that several conceptual problems -- like curvature singularities and ultraviolet divergences -- could be solved. However, in quantum general relativity $\ell_{\rm p}$ and $m_{\rm p}$  are \textit{not} physical cut-offs. Moreover, the above arguments only apply outside the black holes; in fact, when approaching very short distances inside a black hole, the accelerations may become larger. At those scales, to even address any question one would require a consistent and predictive (possibly four-dimensional) theory of quantum gravity that can naturally incorporate the above concepts, e.g. the notion of minimal length or maximal acceleration.

Many quantum-gravity approaches take the existence of a fundamental scale as a starting assumption, and only concern modifications and/or extensions of the low-energy dynamics without paying much attention to kinematics. 
In this Essay we wish to take a different viewpoint which is usually underestimated or less appreciated, and that is based on a critical inspection of fundamental principles, and on a possible generalization of the standard relativistic laws of kinematics, which then will also be followed by a revision of the dynamical laws (i.e. Lagrangians and field equations). This logical way of thinking should be seen as a first necessary step towards the formulation of a consistent theory of quantum gravity.

\section*{Maximal Acceleration}

Although we can agree that a physical cut-off scale should exist in Nature, it is \textit{not} clear what dimensionful scale is more fundamental; could it be a length, a time or an acceleration,...?

Too many guesses can often lead on the wrong path. It is wiser to rely on what we already know and do a deeper investigation of standard theories by looking at things from a different angle. This viewpoint was taken by Caianiello~\cite{Caianiello:1981jq,Caianiello-HUP} who was the first to understand that simply by combining notions of special relativity and quantum mechanics (without gravity) one can derive an upper limit to the proper acceleration of any massive particle.

The starting point in Caianiello's poof is the following Heisenberg uncertainty relation~\cite{Caianiello-HUP,Wood:1989qu}:
\begin{equation}
\Delta E\, \Delta v(t)\geq \frac{\hbar}{2}\left|\frac{{\rm d}v(t)}{{\rm d}t}\right|\,,\label{uncert-eq}
\end{equation}
where $\Delta E$ and $\Delta v(t)$ are the uncertainties on energy $E$ and velocity $v(t)$ of a quantum particle of rest mass $m.$ By using the fact that $\Delta v=(\langle v^2 \rangle - \langle v\rangle^2)^{1/2}\leq c,$ and $\Delta E\leq E\leq mc^2$ (in the instantaneous particle rest frame), we obtain the remarkable bound~\cite{Caianiello-HUP}
\begin{equation}
a\leq \frac{2mc^3}{\hbar}\equiv a_{\rm max}\,,\label{Caianiello-max}
\end{equation}
where $a={\rm d}v/{\rm dt}$ is the proper acceleration of the particle and $a_{\rm max}$ is known as \textit{Caianiello's maximal acceleration}. Obviously, such a bound is intrinsically quantum and relativistic, indeed in either limits $c\rightarrow \infty$ or $\hbar\rightarrow 0$ we recover the classical scenario with no limitation, i.e. $a\leq \infty.$

The bound~\eqref{Caianiello-max} does \textit{not} rely on any gravitational feature. However, as mentioned above, in a quantum-gravitational regime we would expect $m\leq m_{\rm p}$ for any elementary particle. Therefore, when gravitational effects are taken into account one has
\begin{equation}
m\leq m_{\rm p}\quad \Rightarrow\quad a\leq \frac{2m_{\rm p}c^3}{\hbar}= 2 a_{\rm p}\,,\label{fundamental-accel}
\end{equation}
which would represent a \textit{fundamental} upper bound on the acceleration of any elementary massive particle, and coincides (up to a factor $2$) with the Planck acceleration in Eq.~\eqref{Planck-acc}. Caianiello's relation~\eqref{Caianiello-max} also implies an upper bound on the force that any elementary particle of mass $m$ can feel, $F\leq 2m^2c^3/\hbar\leq 2m_{\rm p}^2c^3/\hbar= 2c^4/G.$ The last part of the inequality coincides up to a factor of order one with the so called principle of maximum tension postulated by Gibbons~\cite{Gibbons:2002iv}.

Hence, in the quantum-gravitational regime a physical cut-off with dimensions of acceleration (or force) is expected to emerge. This may suggest the more urgent need to find a theory that naturally incorporates such a new scale, even before bothering about a quantum theory of gravity.

\section*{Reciprocity Principle} 

Any new progress in physics should always be supported by some guiding principle. We should indeed ask -- is there any fundamental principle that can be advocated in order to naturally incorporate a new fundamental scale in our theory? The answer is YES!

In 1938, Born realized that physical laws may be invariant under the so called reciprocity\footnote{The word “reciprocity" was chosen in analogy with the “reciprocal lattice ” in the theory of crystals, where the motion of the particle can be mapped from the $x$-coordinate space to the $p$-momentum space~\cite{Born0}.}  transformation~\cite{Born0}
\begin{equation}
x^\mu \rightarrow \alpha\, p^\mu\quad {\rm and}\quad p^{\mu}\rightarrow -\frac{1}{\alpha}\,x^\mu\,,\label{reciprocity-transf}
\end{equation}
where the positive constant $\alpha$ has dimensions $[{\rm kg}^{-1}\,{\rm s}]$ or $[{\rm m}\,{\rm s}^{-1}\,{\rm N}^{-1}].$ Some physical equations that are manifestly invariant under~\eqref{reciprocity-transf} are the Hamilton equations, the commutation relations, the angular momentum components etc. Born promoted the invariance of physical laws under~\eqref{reciprocity-transf} to a symmetry principle known as \textit{reciprocity principle}. Initially, he aimed at constructing a new theory that could predict the meson mass spectrum~\cite{Born}, but his results were invalidated by experiments. However, the validity of the reciprocity principle can not be discarded yet~\cite{Morgan}.

The first main observation made by Born was that the four-dimensional line element  ${\rm d}s^2=\eta_{\mu\nu}{\rm d}x^\mu{\rm d}x^\nu,$ with $\eta_{\mu\nu}={\rm diag}(-1,1,1,1),$ is \textit{not} invariant under~\eqref{reciprocity-transf}, and that this was an important issue to solve even before trying to combine quantum mechanics and general relativity. Thus, he proposed the following generalized eight-dimensional line element which is invariant under~\eqref{reciprocity-transf}:
\begin{equation}
{\rm d}s^2=\eta_{\mu\nu}{\rm d}x^\mu{\rm d}x^\nu+\alpha^2\eta_{\mu\nu}{\rm d}p^\mu{\rm d}p^\nu\,,\quad \alpha=\frac{c^4}{\mathcal{A}^2\hbar}\,,\,\,\label{8-line-element}
\end{equation}
where $\mathcal{A}$ is a fundamental acceleration scale and it represents a maximal value on the acceleration of any massive particles moving along their worldline; this can be easily shown by using $|{\rm d}\vec{x}/{\rm d}t|\leq c$ and the timelike condition ${\rm d}s^2<0$~\cite{Caianiello:1981jq}. Note that, we could have written $\alpha$ in terms of a new fundamental length, mass or force scale too, but we have chosen to treat acceleration (i.e. kinematics) as more fundamental in view of our arguments above. 

We would expect to recover the four-dimensional line element not only in the limit of no upper bound ($\mathcal{A}\rightarrow\infty$) but also when $c\rightarrow \infty$ or $\hbar\rightarrow 0.$  This implies that $\mathcal{A}$ itself must have a certain dependence on $c$ and $\hbar$ in order to recover the correct ``classical'' limit, thus suggesting a proportionality relation between $\mathcal{A}$ and the Planck acceleration $a_{\rm p}$ introduced in Eq.~\eqref{Planck-acc}. 

Hence, we have shown that the reciprocity principle implies the existence of a  fundamental acceleration scale which corresponds to a maximal limiting value on any acceleration in Nature, consistently with the aforementioned concepts pioneered by Caianiello in~\cite{Caianiello:1981jq,Caianiello-HUP}. 
It is worth mentioning that Caianiello considered a similar line element in his geometrical formulation of quantum mechanics~\cite{quantum-geometry,Torrome:2018zck} but he chose the coordinates $\big(x^\mu,\frac{c^2}{\mathcal{A}}\frac{{\rm d}x^\mu}{{\rm d}s} \big).$

\section*{Nonlocality}

The reciprocity principle has far-reaching unexpected consequences, among which the most interesting is that point-like interactions become intrinsically \textit{nonlocal}. Born applied the principle to quantum electrodynamics and showed that the field equations (in Lorenz gauge)  for the potential $A_\mu$ sourced by $J_\mu$ are modified as follows~\cite{Born,Born-nonlocal}
\begin{equation}
e^{-\ell^2\Box}\Box A_\mu =4\pi J_\mu\,,\quad \ell=\frac{c^2}{\mathcal{A}}\,, \label{qed-nonlocal}
\end{equation}
where $\Box=\eta^{\mu\nu}\partial_\mu\partial_\nu$ is the flat d'Alembertian and $\ell$ is a length scale at which nonlocal effects become relevant. The infinite-order derivatives in $e^{-\ell^2\Box}$ can regularize the singularity of Coulomb potential at $r=0$~\cite{Born}. 

The emergence of nonlocality\footnote{Up to our knowledge, this type of nonlocality and non-polynomial differential operator were introduced independently of reciprocity for the first time in 1931 by Born himself~\cite{Born-rumer}.} is unavoidable if the reciprocity principle is valid, as clearly stated by Born in~\cite{Born}: \textit{``...in the reciprocity theory it is a necessity.''} 
Nonlocal field theories were further investigated by Pais~\cite{Pais:1950za}, Efimov~\cite{Efimov:1967pjn}, and in a slightly different version by Yukawa~\cite{Yukawa:1950eq} . 

We now wish to go beyond Born's pioneering work, and argue that the same nonlocal modification happens if we apply the reciprocity principle to gravity~\cite{progress}.  The gravitational analogue of Eq.~\eqref{qed-nonlocal} is given by the linearized field equations for the metric perturbation around Minkowski, $h_{\mu\nu}=g_{\mu\nu}-\eta_{\mu\nu},$ written in de Donder gauge:
\begin{equation}
e^{-\ell^2\Box}\Box\left(h_{\mu\nu}-\frac{1}{2}\eta_{\mu\nu}h\right) =-16\pi G\, T_{\mu\nu}\,, \label{grav-nonlocal}
\end{equation}
where $h=\eta^{\mu\nu}h_{\mu\nu}$ and $T_{\mu\nu}$ is the stress-energy tensor of matter. Remarkably, exactly the same type of ``nonlocal gravity" has been investigated intensively in the last decades in relation to the problems of renormalizability, unitarity and singularities in quantum gravity~\cite{Krasnikov,Biswas:2005qr}.

Analogously to the case of Coulomb's potential, nonlocality modifies the short-distance behavior of Newton's potential which, in presence of a point-like source $T_{\mu\nu}=M\delta_{\mu}^0\delta_\nu^0\delta^{(3)}(\vec{r}),$ reads
\begin{equation}
\Phi(r)=-\frac{1}{2}h_{00}(r)=-\frac{GM}{r}{\rm Erf}\left(\frac{r}{2\ell}\right)\,,\label{potential}
\end{equation}
and it is indeed regular at the origin, $\Phi(r=0)=-GM/(\ell \sqrt{\pi}).$ 

Furthermore, one can show that the minimal form of the gravitational action compatible with the field equations~\eqref{grav-nonlocal}, and therefore with the reciprocity principle, must be quadratic in the curvature invariants, and it is given by
\begin{equation}
S=\frac{1}{16\pi G}\int {\rm d}^4x\sqrt{-g}\left(R+G_{\mu\nu}\frac{e^{-\ell^2\Box}-1}{\Box}R^{\mu\nu}\right)\,, \label{grav-action}
\end{equation}
where $R_{\mu\nu}$ is the Ricci tensor, $R$ the Ricci scalar and $G_{\mu\nu}$ the Einstein tensor.

\begin{figure}[t]
	\includegraphics[scale=0.41]{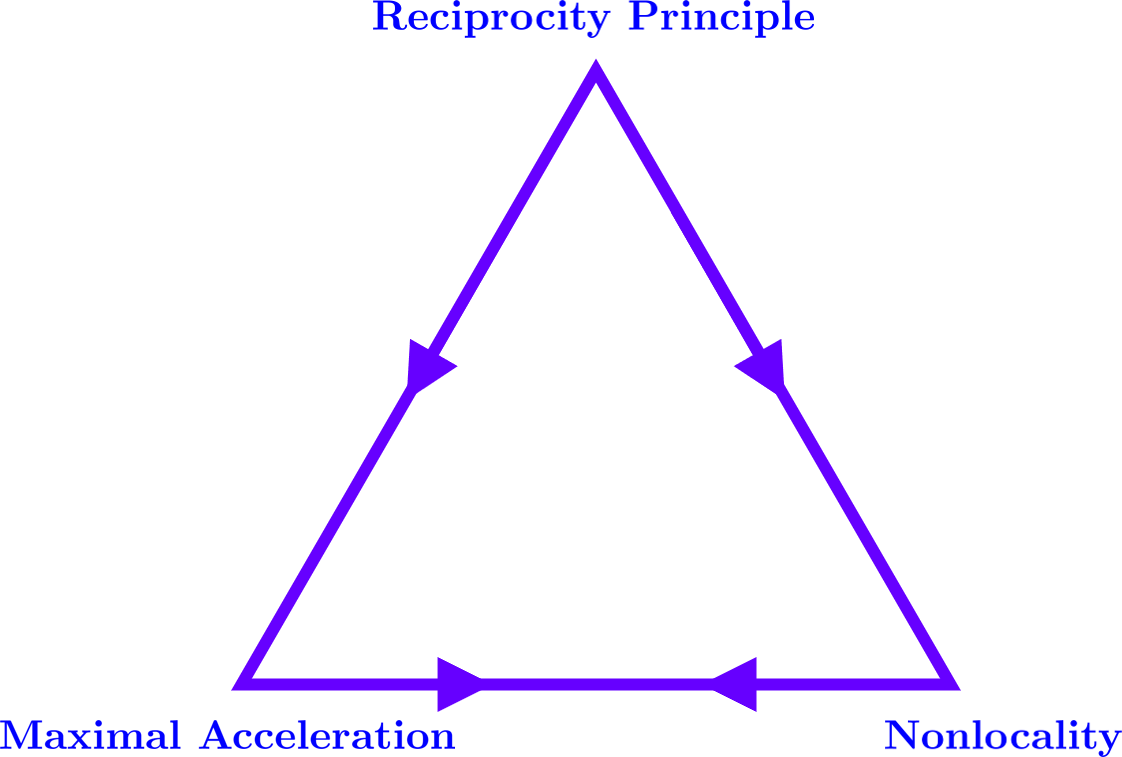}
	\centering
	\protect\caption{Schematic illustration summarizing the intriguing relations among reciprocity principle, maximal acceleration and nonlocality. 
	}\label{fig1}
\end{figure}

\section*{Remarks \& Outlook}

In Fig.~\ref{fig1} we have schematically summarized the content of this work. Let us now make some final important remarks.
\begin{itemize}
	
	\item The scale $\mathcal{A}$ represents a fundamental limit on the  acceleration. If the weak equivalence principle is valid, then an upper bound on the gravitational acceleration (or force) must exist. This feature may be related to the fact that the gravitational potential~\eqref{grav-nonlocal} is bounded. 
	Similar singularity avoidances due to a maximal acceleration were also found in Caianiello's theory~\cite{Feoli:1999cn} and Loop Quantum Gravity~\cite{Rovelli:2013osa}. 
	
	\item The fundamental scale must be an invariant, i.e. the same for any observer, therefore a revision of the transformation laws in relativistic kinematics may be required. A generalized transformations' group that leaves invariant both an eight-dimensional line element and a maximal acceleration was found in~\cite{Scarpetta:1984nn}.
	
	\item A maximal acceleration implies that any would-be point-like object is instead extended~\cite{Caianiello:1981is}. The same interplay characterizes string theory, where the extended nature of strings implies a maximal acceleration~\cite{Frolov:1990ct}. Other features shared with string theory are the nonlocal operator $e^{-\ell^2\Box}$~\cite{Sen:2004nf,Tseytlin:1995uq}, and a connection between reciprocity and T-duality~\cite{Veneziano:1986zf}.
	
	\item Nonlocal theories of gravity~\cite{Krasnikov,Biswas:2005qr} are often criticized for being too arbitrary and \textit{not} motivated from first principles. On the contrary, we argued, for the first time, that the reciprocity principle implies that gravity is fundamentally nonlocal, and uniquely selects one single form of nonlocality given by the exponential in Eqs.~\eqref{grav-nonlocal} and~\eqref{grav-action}.

\end{itemize}

In this Essay we pointed out that quantum-relativistic arguments suggest a critical revision of first principles and standard kinematic laws to incorporate a new fundamental acceleration scale. The reciprocity principle can naturally do so, and as a consequence a nonlocal generalization of dynamical laws emerges. 
Future investigations are necessary to inspect the validity of the reciprocity principle and the existence of the maximal bound on acceleration, especially in relation to the formulation of a consistent quantum theory of gravity~\cite{progress}.


\subsection*{Acknowledgements}
I dedicate this work to the memory of Eduardo~R.~Caianiello whom I was not lucky enough to meet.  I acknowledge financial support from JSPS and KAKENHI Grant-in-Aid for Scientific Research No.~JP19F19324.



\end{document}